\pdfoutput=1
\newcommand*{\ATLASLATEXPATH}{}

\pdfoutput=1
\documentclass[cernpreprint, texlive=2016, USenglish, LANGEDIT=true]{\ATLASLATEXPATH atlasdoc}

\pdfoutput=1

\usepackage{\ATLASLATEXPATH atlaspackage}
\usepackage{\ATLASLATEXPATH atlasbiblatex}

\usepackage{\ATLASLATEXPATH atlasphysics}

\usepackage[centering,scale=0.75]{geometry}
\usepackage{graphicx}
\usepackage{multirow}

\graphicspath{{}{}}

\usepackage{HIGG-2016-16-PAPER-defs}

\AtlasTitle{Search for Resonant and Non-Resonant Higgs Boson Pair Production in the {\boldmath${b\bar{b}\tau^+\tau^-}$} Decay Channel in $pp$ Collisions at $\sqrt{s}$~=~13\,\TeV\ with the ATLAS detector}

\author{The ATLAS Collaboration}

\date{\today}

\AtlasRefCode{HIGG-2016-16}

\PreprintIdNumber{CERN-EP-2018-164}

\AtlasJournalRef{Phys. Rev. Lett. 121, 191801 (2018)}
\AtlasDOI{10.1103/PhysRevLett.121.191801}

\AtlasAbstract{%
A search for resonant and non-resonant pair production of Higgs bosons in the
$b\bar{b}\tau^+\tau^-$~final state is presented. The search uses 36.1 \invfb~of $pp$ collision data
with $\sqrt{s}$~= 13 TeV recorded by the ATLAS experiment at the LHC in 2015 and 2016. Decays of the $\tau$-lepton pairs with at least one $\tau$-lepton decaying to final states with hadrons and a neutrino are considered. No significant
excess above the expected background is observed in the data. The cross-section times branching
ratio for non-resonant Higgs boson pair production is constrained to be less than 30.9~fb, 12.7
times the Standard Model expectation, at 95\% confidence level. The data are also analyzed to probe
resonant Higgs boson pair production, constraining a model with an extended Higgs sector based on
two doublets and a Randall--Sundrum bulk graviton model. Upper limits are placed on the resonant
Higgs boson pair production cross-section times branching ratio, excluding resonances $X$ in the
mass range 305~\GeV\ < $m_X$ < 402~\GeV\ in the simplified hMSSM minimal supersymmetric model for
$\tan\beta=2$ and excluding bulk Randall--Sundrum gravitons $G_{\mathrm{KK}}$ in the mass range
325~\GeV\ < $m_{G_{\mathrm{KK}}}$ < 885~\GeV\ for $k/\overline{M}_{\mathrm{Pl}} = 1$.
}

\hypersetup{pdftitle={ATLAS},pdfauthor={The ATLAS Collaboration}}
\DeclareUnicodeCharacter{2212}{-}
\DeclareUnicodeCharacter{2009}{ }
\begin{document}
\newif\ifhadhadnote

\maketitle




In 2012, the ATLAS and CMS Collaborations at the LHC discovered a new particle with a mass of
approximately 125~\GeV\ \cite{HIGG-2012-27, CMS-HIG-12-036, Higgs_mass}. According to all current
measurements it is compatible with the Standard Model (SM) Higgs boson ($H$) \cite{Englert_Brout, Higgs, Guralnik, Higgs:1966, Kibble}. 
An important pending
test of the Brout--Englert--Higgs mechanism is the measurement of Higgs boson pair production.  At the LHC, pairs of
SM Higgs bosons can be produced via the Higgs self-interaction (`triangle-diagram') and the destructively interfering 
top-quark loop (`box-diagram')~\cite{HH_xsec,deFlorian:2016spz}. Non-resonant Higgs
boson pair production (NR $HH$) can be significantly enhanced relative to the SM prediction by modifications
to the top-quark Yukawa coupling, the trilinear Higgs boson coupling $\lambda_{HHH}$ or by
introducing production mechanisms with new intermediate particles. Many theories beyond the SM predict
heavy resonances that could decay into a pair of SM Higgs bosons, such as a heavy $CP$-even scalar $X$
in two-Higgs-doublet models \cite{Branco:2011iw}, or spin-2 Kaluza--Klein (KK) excitations of
the graviton, $G_\mathrm{KK}$, in the bulk Randall--Sundrum (RS) model
\cite{Randall:1999ee,Agashe:2007zd,Fitzpatrick:2007qr}.

This Letter describes a search for resonant and non-resonant Higgs boson pair production in a final
state with two $b$-quarks and two $\tau$-leptons using 36.1 $\invfb$~of $pp$ collision data recorded with the ATLAS
detector~\cite{PERF-2007-01, Capeans:2010jnh} in 2015 and 2016. The \tlhad and \thad\thad\ decay channels are considered, where the subscripts (lep = electron or muon,
had = hadrons) indicate the decay mode of the $\tau$-lepton.
Previous searches for Higgs boson pair production were performed at center-of-mass energies
$\sqrt{s}$ = 8 \TeV\ \cite{Aad2015xja, Aad2015uka, Aadyja, Aad2015xja} and $\sqrt{s}$ = 13
\TeV\ \cite{Aaboud:2018knk,Sirunyan:2017djm, Sirunyan:2017isc} by the ATLAS and CMS
Collaborations. The ATLAS search in the $4b$ channel constitutes the most sensitive result to date and the observed (expected) limit 
excludes a cross-section greater than 13.0 (20.7) times the SM prediction at 95\% confidence level (CL).


The SM non-resonant $HH$ process was simulated with \MGMCatNLO at next-to-leading order (NLO)
\cite{MG5AMCATNLO,Frederix:2014hta,Dawson:1998py,Plehn:1996wb,Grober:2015cwa} using the CT10 parton
distribution function (PDF) set~\cite{Lai:2010vv}.  Parton showers and hadronization were simulated
with \Herwigpp \cite{herwigpp} using the UEEE5 set of tuned parameters (tune) \cite{Gieseke:2012ft}.
The events were reweighted to reproduce the $m_{HH}$ spectrum obtained in
Refs.~\cite{HH_xsec,finite_mtop_2}, which fully accounts for the finite mass of the top quark. The
cross-section times branching ratio to the $bb\tau\tau$ final state, evaluated at
next-to-next-to-leading order (NNLO) and including next-to-next-to-leading logarithm (NNLL)
corrections and NLO top-quark mass effects, is $2.44^{+ 0.18}_{-0.22} $~fb~\cite{finite_mtop_1}.
Events with a generic narrow-width scalar $X$ or $G_\mathrm{KK}$ decaying into $HH$ were produced
in \MGMCatNLO at leading order (LO) and interfaced to the \textsc{Pythia~8} \cite{ Pythia8} parton
shower model using the A14 tune \cite{A14tune} together with the NNPDF23LO PDF
set~\cite{Ball:2012cx}.  The cross-section and width of the $G _\mathrm{KK}$ were taken from
Ref.~\cite{Oliveira:2014kla} and depend on $k/\overline{M}_\mathrm{Pl}$, where $k$\ corresponds to
the curvature of the warped extra dimension and $\overline{M}_\mathrm{Pl} = 2.4 \times 10^{18}$
\GeV\ is the effective four-dimensional Planck scale. Events with $k/\overline{M}_\mathrm{Pl}=1$ and
$k/\overline{M}_\mathrm{Pl}=2$ were simulated.

The dominant SM background processes are
\ttbar, QCD multi-jet and $Z$ bosons produced in association with jets originating from heavy-flavor
quarks ($bb, bc, cc$), subsequently referred to as $Z$+heavy-flavor~\cite{lightflavor}.
SM Higgs boson production in association with a $Z$\ boson,
subsequently decaying into a $bb\tau\tau$~\cite{tautau}
%
%
final state, is an irreducible background in this analysis.
The \ttbar\ and single-top-quark background events were simulated using
\textsc{Powheg-Box}~\cite{Powheg3}, with the CT10 PDF set, and \textsc{MadSpin}~\cite{MadSpin}. The
parton showers were simulated using \textsc{Pythia~6}~\cite{Pythia} and the Perugia 2012 tune
\cite{Perugia}.  The \ttbar\ background was scaled to match the NNLO+NNLL
cross-sections~\cite{ttbar_xsec}, while the single-top samples were corrected to
NLO~\cite{Aliev:2010zk,Kant:2014oha} (approximate NNLO~\cite{Kidonakis:2010ux}) predictions for the
$t$- and $s$-channel ($Wt$ final state).  Events with $W$\ or $Z$\ bosons and associated jets were
simulated with the \textsc{Sherpa 2.2.1} generator \cite{SHERPA, Comix, OpenLoops, CSShower,
  MEPS@NLO}, using the NNPDF30NNLO PDF set~\cite{Ball:2014uwa} and normalized to the NNLO cross-sections
~\cite{WZcrossSections2}.  Diboson and Drell--Yan backgrounds were produced with
\textsc{Sherpa~2.1.1}~\cite{SHERPA} using the CT10NLO PDF set and the generator cross-section
predictions.  Quark-induced $ZH$\ processes were generated with \textsc{Pythia~8}, using the A14
tune and the NNPDF23LO PDF set. The samples were normalized to NNLO cross-sections for QCD and NLO
for electroweak
processes~\cite{Ciccolini:2003jy,Brein:2003wg,Ferrera:2011bk,Brein:2011vx,Ferrera:2013yga,Ferrera:2014lca,Campbell:2016jau}. The
gluon-induced $ZH$\ process~\cite{POWHEG} was generated with \textsc{Powheg} using the CT10 PDF set
and using \textsc{Pythia~8} with the AZNLO tune~\cite{AZNLOtune} to simulate parton
showers. Cross-sections~\cite{Altenkamp:2012sx,Hespel:2015zea,ggzhnll,Harlander:2013mla,Brein:2012ne}
were scaled to NLO+NLL in QCD. SM Higgs boson production in association with a top-quark pair was
simulated with \MGMCatNLO; \textsc{Pythia~8} was used to simulate the parton shower, while the
cross-section was taken from Ref.~\cite{deFlorian:2016spz}.  In all signal and background samples,
the mass of the $H$ bosons was set to 125 GeV.  The contributions from other SM Higgs boson
processes are negligible. \textsc{EvtGen} v1.2.0~\cite{EvtGen} was used to model the properties of
bottom and charm hadron decays for all processes except those simulated in \textsc{Sherpa}. The
detector response to the generated events was simulated with \textsc{Geant4}~\cite{Geant4,
  ATLASSIM}. Simulated events are reweighted to match the distribution of the number of inelastic
collisions per event (pileup) in data.


Events are required to have at least one collision vertex reconstructed from at least two
charged-particle tracks with transverse momentum~\cite{coordinates}
$p_\mathrm{T}^{\mathrm{track}}>0.4$~GeV. The primary vertex for each
event is selected as the vertex with the highest $\sum(p_\mathrm{T}^{\mathrm{track}})^2$.  Jets are
formed using the anti-$k_t$ algorithm~\cite{Cacciari:2008gp} with a radius parameter $R=0.4$ and
calorimeter energy clusters as inputs~\cite{Aad:2016upy, Aaboud:2017jcu, ATLAS-CONF-2014-018}.
These jets are taken as seeds for the reconstruction of the visible products of hadronically decaying
$\tau$-leptons (\tauhadvis) \cite{ATLAS_run2_taus, ATL-PHYS-PUB-2015-045, Aad:2014rga}, which are
subsequently required to have one or three associated tracks.  In order to distinguish
\tauhadvis\ from quark- and gluon-initiated jets, a boosted decision tree (BDT)~\cite{Hocker:2007ht}, trained separately for
\tauhadvis\ with one and three charged particles, is employed.  Selected \tauhadvis\ candidates must
satisfy the `medium' BDT working point~\cite{ATL-PHYS-PUB-2015-045}.  Electron candidates are identified using a likelihood
technique in combination with additional track-hit requirements \cite{ATLAS:2016iqc};
the transition region between the barrel and endcap calorimeters is excluded. 
Information from the tracking and muon systems is used to reconstruct muon candidates~\cite{Aad:2016jkr}.
Only isolated electrons and muons are considered, where no nearby tracks or calorimeter energy
deposits within a \pt-dependent variable-size $\Delta R$ cone around the lepton are allowed.  Jets
arising from pileup are suppressed using dedicated track and vertex requirements~\cite{JVT}.  The
missing transverse momentum, with magnitude \met, is defined as the negative vectorial sum of all
reconstructed and fully calibrated objects in the event, along with an additional track-based soft
term~\cite{METCP}.  Jets containing $b$-hadrons are identified using the MV2c10 multivariate
discriminant \cite{bTagging_Run1,bTagging_Run2} trained against a light-quark-flavor sample also
containing 10\% of $c$-hadrons. A working point with 70\% efficiency on simulated \ttbar\ events is used. 
An overlap-removal procedure is applied to the reconstructed electrons, muons,
\tauhadvis\ and jets to prevent double-counting of energy deposits in the detector as described in Ref.~\cite{HeavyHtotautau}.

The selected final state is characterized by one electron or muon and one \tauhadvis\ of opposite
charge, or two \tauhadvis\ of opposite charge, plus two $b$-tagged jets and \met. In all cases,
events with additional electrons or muons above 7~\GeV\ or \tauhadvis\ above 20~\GeV\ are rejected.  
The offline selection criteria
for the electron, muon, and \tauhadvis\ depend on the triggers used. In the \tlhad channel events
are selected with a single-lepton trigger (SLT) and a lepton plus \tauhad\ trigger (LTT), which are
analyzed separately and combined with the \thadhad channel in the final fit. Depending on the data
period, the electron or muon that passes the SLT trigger is required to have \pt\ >~25--27~\GeV.
Events which fail this requirement are considered for the LTT category if the electron (muon) has
\pt\ >~18~\GeV\ (15~\GeV). In all cases, these \pt\ requirements are 1 \GeV\ higher than the trigger
thresholds to ensure a nearly constant trigger efficiency relative to the offline selection.  The
\tlhad events are required to have one \tauhadvis\ candidate with $|\eta|$~<~2.3 and
\pt\ >~20~\GeV\ for SLT events, raised to 30~\GeV\ for LTT events due to
\tauhadvis\ \pt\ requirements applied in this category of triggers.  In the \thadhad channel a
logical OR of single \tauh\ triggers (STT) and di-\tauh\ triggers (DTT) is used. The leading
\tauhadvis\ candidate is required to have a minimum \pt\ of 40~\GeV\ for DTT and between 100 and 180~\GeV 
for STT events, depending on the
data-taking period. The sub-leading \tauhadvis\ is required to have a minimum
\pt\ of 20 (30)~\GeV\ for STT (DTT) events.  The leading jet is required to have \pt >~45~\GeV,
except in the LTT and DTT channels where this is raised to 80 GeV due to a requirement on the
presence of a jet at the Level-1 trigger to reduce the rate (during 2016 data-taking only for the
DTT).  In all cases the sub-leading jet must have \pt~>~20~\GeV.  The invariant mass of the
di-$\tau$ system, \mmmc, is calculated using the Missing Mass Calculator \cite{MMC} and is required
to be greater than 60~\GeV.  Signal region (SR) events are defined as those meeting the criteria
above, and in addition containing two $b$-tagged jets; they are further separated into \tlhad SLT,
\tlhad LTT and \thadhad categories.


BDTs are used in the analysis to improve the separation of signal from background.  Their
distributions in the three signal regions, along with control region yields to constrain the
normalization of the dominant backgrounds, form the inputs to the final fit.
The BDTs for the \thadhad channel are trained against the main backgrounds, \ttbar, $Z \to \tau\tau$ and
multi-jet events; in the \tlhad channel they are trained solely against the dominant \ttbar\ background.
For the BDT trainings, the \ttbar\ and $Z \to \tau\tau$ backgrounds are taken purely from simulation, 
while the multi-jet events are estimated using the data-driven approach described below.
Variables which provide good discrimination and are minimally correlated are used as inputs
to the BDTs, as summarized in Table~\ref{tab:bdt_vars}.  The variables selected in each channel
differ, reflecting the different background compositions.  In the resonant search, BDTs are trained separately
for each signal mass considered, from 260~\GeV\ to 1000~\GeV\ (800~\GeV\ for LTT), where
the signal model combines the target resonance mass and its two neighboring mass points, to be
sensitive to masses between the simulated points.  For NR $HH$ production, the BDTs are trained on a
signal sample with the SM admixture of the contributions from the box-diagram and triangle-diagram.  The BDTs are 
more sensitive to the box-diagram where the two Higgs bosons are produced at higher \pt\ and the
selection efficiency is greater.

\begin{table}
\centering
\caption{ Variables used as inputs to the BDTs for the different channels and signal models.  Here,
  \mhh\ is reconstructed from the $\tau\tau$ and $bb$ systems using a 125\,\GeV\ Higgs mass
  constraint; \mmmc is the invariant mass of the di-$\tau$ system, calculated using the Missing Mass Calculator \cite{MMC}; \mbb\ is the invariant $bb$-mass; \dRtautau\ is evaluated
  between the electron or muon and \tauhadvis\ (two \tauhadvis) in the case of the \tlhad
  (\thadhadalt) channel; \met $\phi$ centrality quantifies the relative angular position of the
  \met\ relative to the visible $\tau$ decay products in the transverse plane~\cite{Aad:2015vsa} and
  is defined as $(A + B)/(\sqrt{A^{2} + B^{2}})$, where 
  $A=\sin(\phi_{\met}-\phi_{\tau_{2}})/\sin(\phi_{\tau_{1}}-\phi_{\tau_{2}})$, 
  $B=\sin(\phi_{\tau_{1}}-\phi_{\met})/\sin(\phi_{\tau_{1}}-\phi_{\tau_{2}})$,  and 
  $\tau_{1}$ and $\tau_{2}$ stand for electron or muon and \tauhadvis\ (two \tauhadvis) in the case of the \tlhad
  (\thadhadalt) channel;  \mtw\ is the transverse mass of the lepton and the \met; 
   \dPhiHH\ is the azimuthal angle between the two Higgs boson candidates; \dptltau\ is 
  the difference in \pt\ between the electron or muon and \tauhadvis.
\label{tab:bdt_vars}} 
  \begin{tabular}{  l  c  c  c } \hline\hline
    \multirow{2}{*}{Variable}   & \tlhad channel  & \tlhad channel           &  \multirow{2}{*}{\thadhad channel} \\
                                &  (SLT resonant)  & (SLT non-resonant \& LTT) &                                    \\[0.1cm] \hline
    \mhh                        & \checkmark       & \checkmark                & \checkmark                         \\ [0.1cm]
    \mmmc                       & \checkmark       & \checkmark                & \checkmark                         \\
    \mbb                        & \checkmark       & \checkmark                & \checkmark                         \\
    \dRtautau                   & \checkmark       & \checkmark                & \checkmark                         \\
    \dRbb                       & \checkmark       & \checkmark                & \checkmark                         \\
    \met                        & \checkmark       &                           &                                    \\[0.1cm]
    \met $\phi$ centrality      & \checkmark       &                           & \checkmark                         \\[0.1cm]
    \mtw                        & \checkmark       & \checkmark                &                                    \\
    \dPhiHH                     & \checkmark       &                           &                                    \\
    \dptltau                    & \checkmark       &                           &                                    \\ 
    Sub-leading $b$-jet \pt     & \checkmark       &                           &                                    \\ \hline\hline
  \end{tabular}
\end{table}


In both channels, simulated events are used to model background processes containing reconstructed
\tauhadvis\ that are matched to generated \tauh\ within $\Delta R = 0.2$ (subsequently referred to
as true-\tauh) and other minor background contributions. The rate of events with at least one true-\tauh\ and a jet
reconstructed as an electron or muon is found to be negligible.  For \ttbar\ background events
containing one or more true-\tauh\ the normalization is obtained in the final fit,
constrained mainly by the low \tlhad\ BDT score regions, resulting in a normalization factor of $1.06
\pm 0.13$.  The normalization of the $Z \to ee / \tau\tau$+heavy-flavor background is determined
using $Z\rightarrow\mu\mu$+heavy-flavor events. Their selection closely follows the event
selection used for signal events. Instead of two $\tau$-lepton candidates, two muons with \pt $>$
27~\GeV\ and dimuon invariant mass between 81 and 101~\GeV\ are selected. To remove
the contribution from SM $ZH(H\rightarrow bb)$ production, $m_{bb}$ is required to be lower than
80~\GeV\ or greater than 140~\GeV. The normalization is determined by including the
$Z\rightarrow\mu\mu$+heavy-flavor control region yield in the final fit, resulting in a
normalization factor of $1.34\pm 0.16$. Normalization factors are not applied to the
$Z$+light-flavor contributions. The modeling of the
BDT score distributions is validated in the 0-$b$-tag and 1-$b$-tag regions as well as in dedicated
\ttbar\ and $Z$+heavy-flavor validation regions.

Contributions from processes in which a quark- or gluon-initiated jet is misidentified as a
\tauhadvis\ candidate (fake-\tauh) are estimated using data-driven methods for major backgrounds.  A
fake-\tauh\ enriched sample is defined by requiring that a \tauhadvis\ fails the `medium' BDT
identification but satisfies a very loose requirement on the BDT score. This selection maintains a
composition of quark- and gluon-initiated jets similar to those mimicking \tauhadvis\ in the SR.  In
the case where the event contains more than one such fake-\tauh, one is chosen randomly.  The SR
selection, except for the \tauhadvis\ identification, is applied to the fake-\tauh\ enriched sample to
extract template distributions for the fake-\tauh\ background after the true-\tauh\ contamination is
subtracted using simulation.  The templates are scaled with fake-factors (FF) defined as the ratio
of the number of fake-\tauh\ that pass the $\tauhadvis$ identification to the number that fail,
calculated in dedicated control regions (CR) and parametrized in \pt(\tauhadvis) and
the number of associated tracks.

For the \tlhad final state, fake-\tauh\ background contributions from \ttbar, $W$+jets and multi-jet
processes are estimated using a combined fake-factor method similar to that described in
Ref.~\cite{HeavyHtotautau,ttbarfake}
%
%
In order to account for the different sources of fake-\tauh, the FFs are derived separately for each
background contribution. The CR for multi-jet events is defined by inverting the isolation
requirement applied to the electron or muon for events with 0 or 1 $b$-tagged jet. The
\ttbar\ ($W$+jets) control region is defined by requiring two (zero) $b$-tagged jets and
\mtw~$>~40$~\GeV, where $\mtw = \sqrt{2 \pt^{\mathrm{lep}} \met (1 -\cos{\Delta\phi_{\mathrm{lep},\met} })}$, and 
$\Delta\phi_{\mathrm{lep},\met}$ is the azimuthal angle between the electron or muon and the \met.  Fake-factors for \ttbar\ and $W$+jets are found to be
consistent for both processes.  The individual fake-factors are then combined as
$\mathrm{FF}(\mathrm{comb}) = \mathrm{FF}(\mathrm{QCD}) \times r_{\mathrm{QCD}} +
\mathrm{FF}(\ttbar/W+\mathrm{jets}) \times (1-r_{\mathrm{QCD}})$, where $r_{\mathrm{QCD}}$ is
defined as the fraction of fake-\tauh\ from (predominantly multi-jet) processes contributing to
the data in the fake-\tauh\ enriched template region that are not accounted for by simulated
background processes, and is less than $5\%$ in the 2-$b$-tag region.  Because of the different
origin of fake-\tauh, the FFs for \ttbar/$W$+jets can be up to 30\% larger than those for
multi-jet processes. Events with two $b$-tagged jets but a same-sign charge (SS) electron or muon
and \tauhadvis\ are used for validating the fake-\tauh\ background, showing all distributions are
well modeled.  Given this, and the small size of the contribution, no transfer factor is applied to
correct the multi-jet estimation from the 1-$b$-tag region to the 2-$b$-tag region.

In the \tauh\tauh\ final state, only the multi-jet background is estimated from data using the FF
method.  The differential FFs are derived in a 1-$b$-tag SS control region, while the overall
normalization is taken from the 2-$b$-tag SS control region.  The \ttbar\ background is
estimated from simulation, where the fake-$\tauh$ \ttbar\ contribution is corrected in bins
of $\eta(\tauhadvis)$ using the probability for a jet from a hadronic $W$-boson decay to mimic a
\tauhadvis\ candidate (fake-rate), as measured with data in the \tlhad\ \ttbar\ control
region~\cite{HeavyHtotautau}.  Contributions from true-\tauh\ are subtracted using simulation.  


The uncertainty in the integrated luminosity of the combined 2015+2016 dataset is 2.1\%~\cite{lumi}
and is applied to the signal and background components whose normalizations are derived from
simulation.  An uncertainty related to the pileup reweighting procedure is also
applied~\cite{PileupCP}.
Experimental uncertainties in the identification and reconstruction of the electron~\cite{ElecCP},
muon~\cite{MuonCP}, \tauhadvis~\cite{ATLAS_run2_taus}, and jets~\cite{JetCP1,Aaboud:2017jcu} are
accounted for and propagated through the analysis to determine their effect on the final results.
These affect the trigger requirements, the identification and reconstruction efficiencies, the
isolation, and the reconstructed energies and their resolutions.  The uncertainties
are propagated to the calculation of the \met~\cite{METCP}, which has an additional uncertainty from
the soft term.  
The uncertainties with the largest impact on the result are those related to the \tauhadvis\, 
identification efficiency, which correspond to an uncertainty of 16\% on the NR signal strength, 
i.e.\ the simulated NR $HH$ yield assuming a cross-section times branching fraction equal
to the expected limit and normalized to the SM expectation ($\sigma^{\mathrm{exp}}/\sigma^{\mathrm{SM}}$).
Uncertainties in flavor-tagging~\cite{BTagCP1,BTagCP2} also have a significant impact, inducing an uncertainty in the
NR signal strength of 8.3\%, dominated by those associated with the $b$-tagging efficiency.

Theory uncertainties in the modeling of the \ttbar\ background containing one or more
true-\tauh\ are assessed by varying the matrix element generator (using \AMCatNLO instead of
\textsc{Powheg-Box}) and the parton shower model (using \Herwigpp instead of \textsc{Pythia~6}), and
by adjusting the factorization and renormalization scales along with the amount of additional
radiation.  The resulting variations in the BDT distributions are included as shape uncertainties in
the final fit.  In order to account for potential acceptance differences between control and signal regions, the
normalization of the \ttbar\ background containing true-\tauh, determined predominantly from the
\tlhad SR in the final fit, is allowed to vary within a range determined by the acceptance
variations associated with the \ttbar\ modeling uncertainties.  This amounts to +30/$-32$\% for the \thadhad
SR and and +8.1\%/$-9.3$\% for the $Z\rightarrow\mu\mu$+heavy-flavor control region.  This is the
dominant uncertainty in the \ttbar\ modeling.

For the $Z$+jets background, the theory uncertainties in the modeling of the BDT shapes are derived
by comparing the nominal \textsc{Sherpa} sample with an alternative \MGMCatNLO+\textsc{~Pythia~8}
sample and by varying the choice of renormalization and factorization scales, along with the PDF
prescription~\cite{PDF4LHC15}. 
The normalization of the $Z \to \tau\tau$+heavy-flavor background in
the \tlhad (\thadhadalt) SR is allowed to vary by 29\% (35\%) 
relative to the normalization derived in the $Z\rightarrow\mu\mu$+heavy-flavor 
control region in order to account for acceptance differences between the two.
An additional 20\% normalization
uncertainty in the $Z \to ee$+light-flavor background, related to the misidentification of electrons 
as taus, is derived by comparing data and simulation in
a $Z\rightarrow ee$ control region with 0 or 1 $b$-tagged jets. 
The $ZH$ ($ttH$) background
normalization is varied by 28\% (30\%) based on ATLAS
measurements~\cite{Aaboud:2017xsd,Aaboud:2017jvq}. 
The normalizations of the remaining minor
backgrounds taken from simulation are allowed to vary within their respective cross-section
uncertainties.

The uncertainty in the modeling of backgrounds due to jets being misidentified as \tauhadvis\ is
estimated by varying the fake-factors and fake-rates within their statistical uncertainties and
varying the amount of true-\tauh\ background subtracted. Based on studies with simulated \ttbar\ and
$W$+jets events, a systematic uncertainty is assigned to cover the difference in the gluon and quark
flavor composition of jets misidentified as a \tauhadvis\ between the signal region and the
fake-\tauh\ enriched sample, parametrized as a function of the \tauhadvis\ identification BDT
score. The uncertainty in the extrapolation of $\mathrm{FF}(\mathrm{QCD})$ to the signal region is
estimated from the difference between the nominal FFs and alternative ones, calculated either in the
SS region for the \tlhad channel or a multi-jet enriched region, where $\Delta\phi(\tauhadvis,\tauhadvis) >
2.0$, in the \thadhad case. Similarly, changes in the fake-\tauh\ determination when varying
the \ttbar\ control region \mtw requirement in simulation and data are used to estimate a systematic
uncertainty in both the fake-factors and fake-rates. The overall effect of these uncertainties on
the fake-\tauh\ background estimate leads to an 8.4\% variation of the NR signal strength, predominantly
due to the true-\tauh\ subtraction in the \ttbar\ control region and the composition of the
fake-\tauh.

Theory uncertainties in the signal acceptance are calculated by independently varying the
renormalization and factorization scales, the choice of PDF and each PDF set by its
uncertainties. The uncertainty in the parton shower is taken into account by comparing the default
\Herwigpp with \textsc{Pythia~8}. Uncertainties in the underlying event, initial-state radiation and
final-state radiation are accounted for by changing the \textsc{Pythia} tune, but are small.
The effects of various categories of uncertainty on the measured non-resonant signal strength
corresponding to the expected upper limit at 95\% CL are summarized in Table~\ref{tab:syst}. The
individual sources of uncertainty making up the categories listed in the table are grouped together
in the final fit to determine their correlated combined effect on the signal strength.   For all
signal hypotheses, the statistical uncertainties dominate.

\begin{table}
\begin{minipage}{0.65\textwidth}
  \begin{tabular}{
      ll
      S[table-format=1(3),
        table-number-alignment=right]
      }
\hline\hline
Source                     &\qquad& \text{Uncertainty (\%)}  \\
\hline
Total                      &\qquad& \text{$\pm$} 54 \\
Data statistics            &\qquad& \text{$\pm$} 44 \\
Simulation statistics      &\qquad& \text{$\pm$} 16 \\
\hline\hline
\multicolumn{2}{c}{Experimental Uncertainties} \\
\hline
Luminosity                 &\qquad& \text{$\pm$} 2.4 \\
Pileup reweighting        &\qquad& \text{$\pm$} 1.7 \\
\tauh\                     &\qquad& \text{$\pm$} 16 \\
Fake-$\tau$ estimation     &\qquad& \text{$\pm$} 8.4 \\
$b$-tagging                &\qquad& \text{$\pm$} 8.3 \\
Jets and $E_\mathrm{T}^{\mathrm{miss}}$&\qquad& \text{$\pm$} 3.3 \\
Electron and muon          &\qquad& \text{$\pm$} 0.5 \\
\hline\hline
\multicolumn{2}{c}{Theoretical and Modeling Uncertainties}\\
\hline
Top                &\qquad& \text{$\pm$} 17 \\
Signal             &\qquad& \text{$\pm$} 9.3 \\
$Z\to\tau\tau$     &\qquad& \text{$\pm$} 6.8 \\
SM Higgs           &\qquad& \text{$\pm$} 2.9 \\
Other backgrounds  &\qquad& \text{$\pm$} 0.3 \\
\hline\hline
\end{tabular}
\end{minipage}
\begin{minipage}{0.33\textwidth}
  \caption{The percentage uncertainties on the simulated non-resonant signal strength, 
i.e.\@\xspace the simulated NR $HH$ yield assuming a cross-section times branching
fraction equal to the 95\% CL expected limit of 14.8 times the SM expectation.}
  \label{tab:syst}
\end{minipage}
\end{table}

\begin{figure}
  \centering 
  \subfloat[]{
    \includegraphics[width=0.35\textwidth]{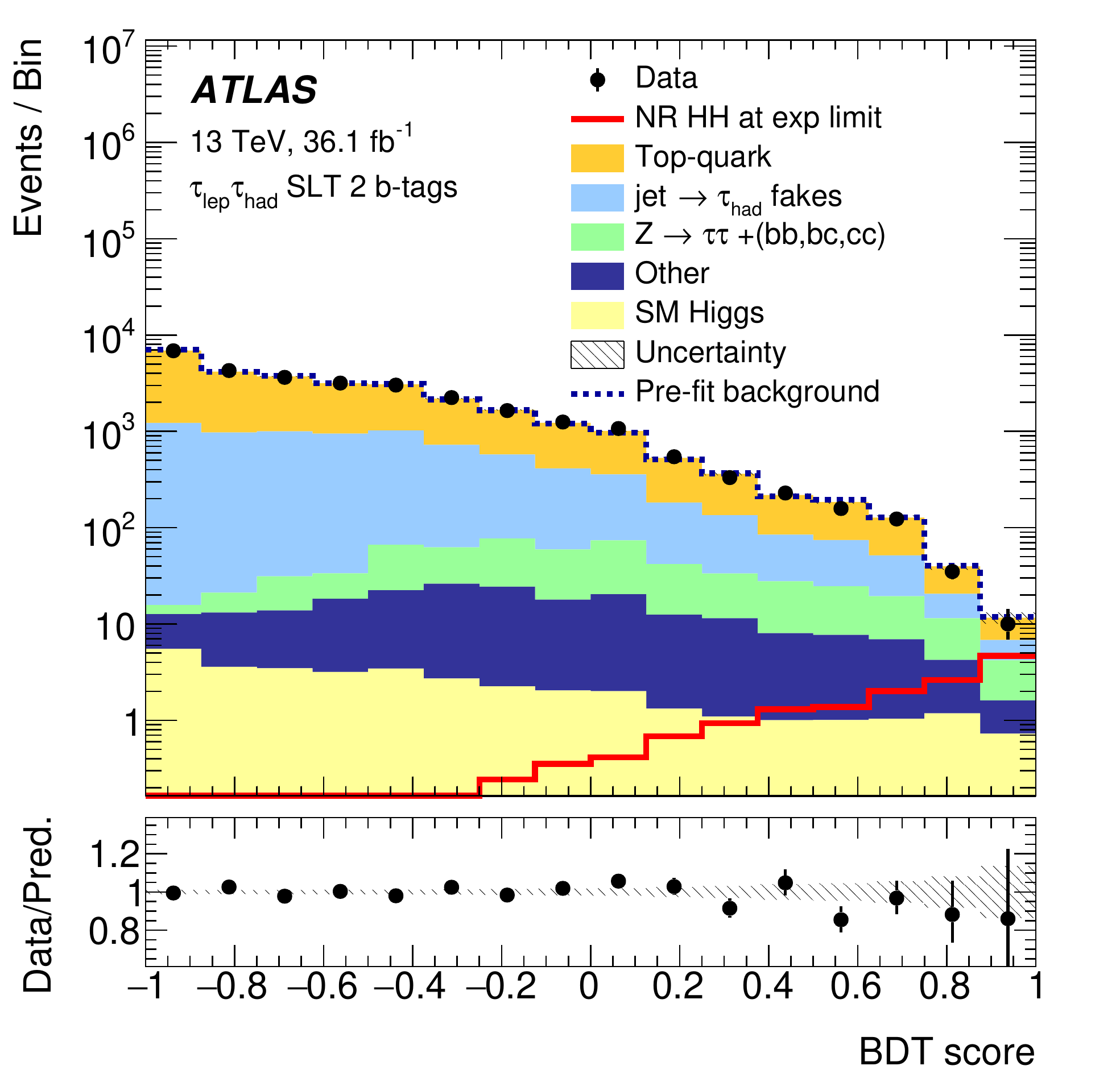}
  }
  \subfloat[]{
    \includegraphics[width=0.35\textwidth]{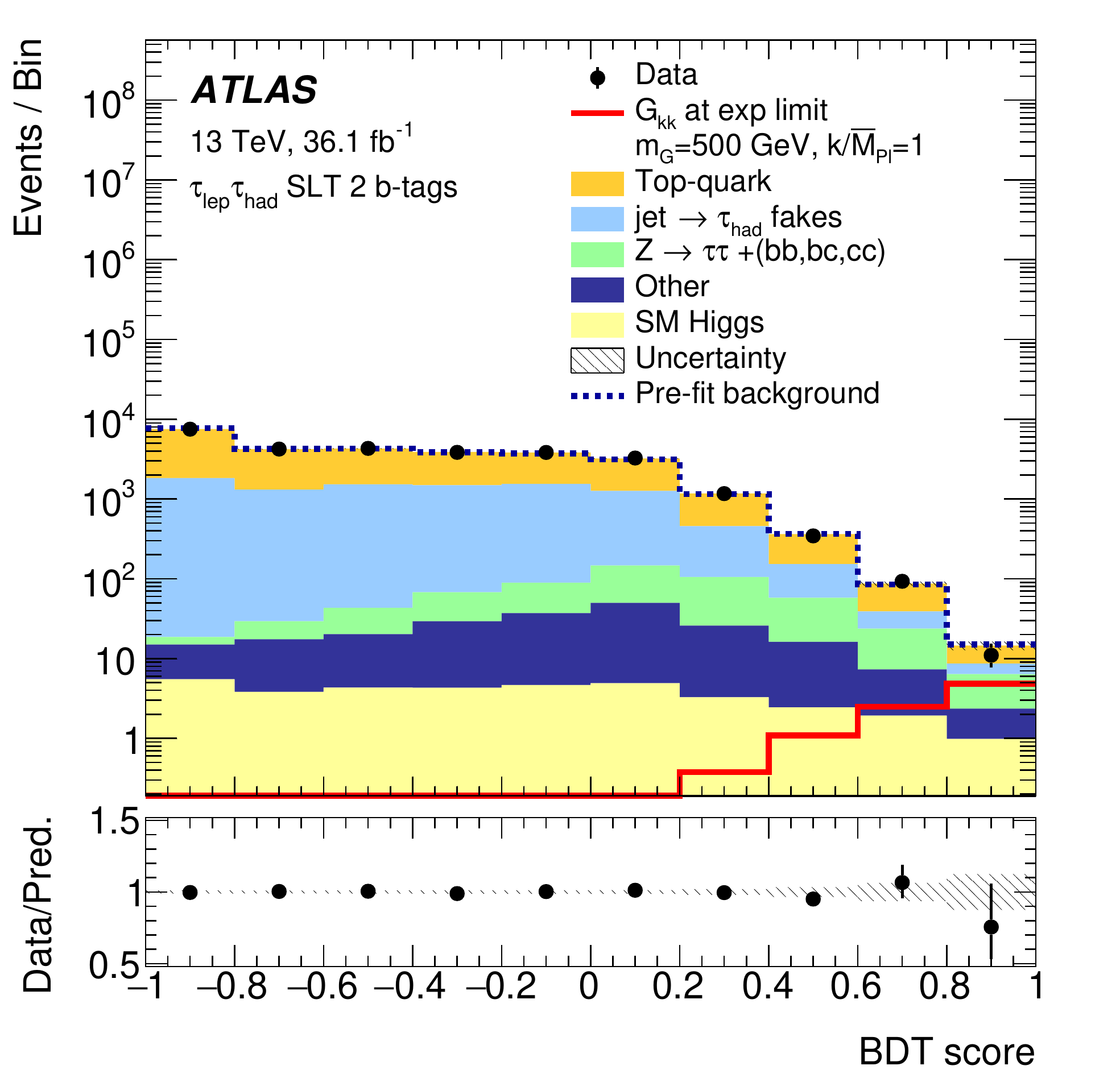}
  }\\
  \subfloat[]{
    \includegraphics[width=0.35\textwidth]{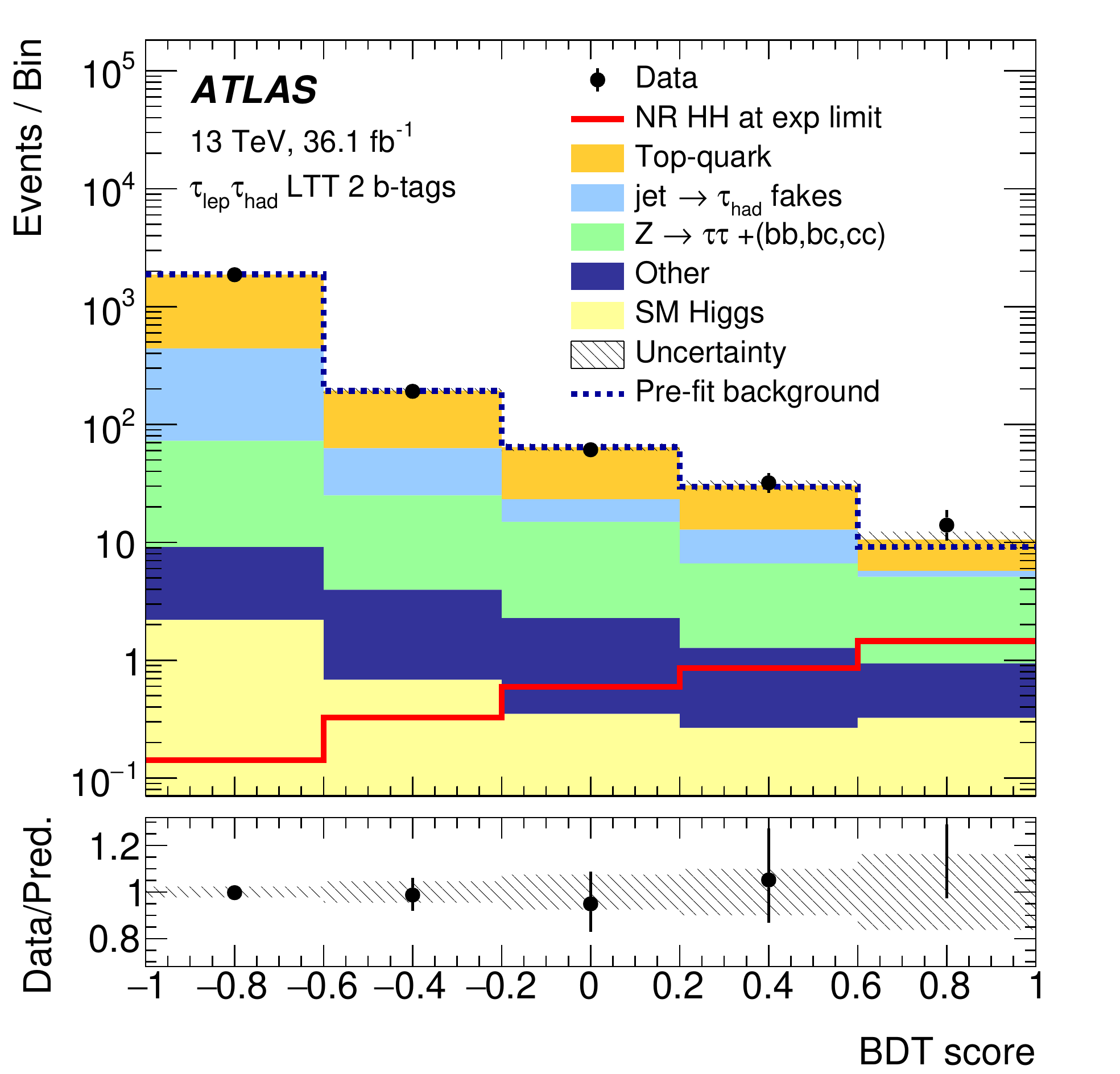}
  }
  \subfloat[]{
    \includegraphics[width=0.35\textwidth]{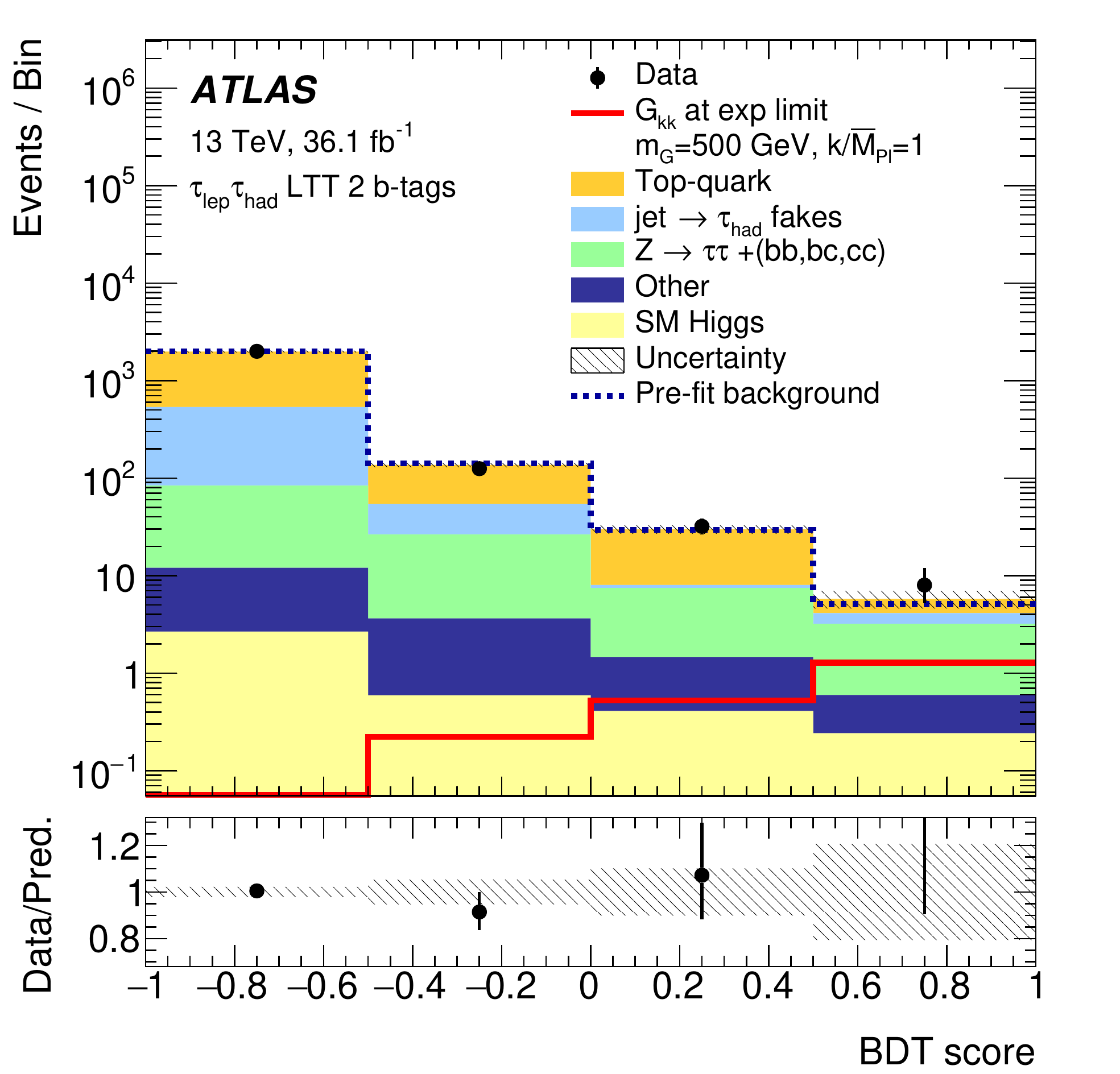}
  }\\
  \subfloat[]{
    \includegraphics[width=0.35\textwidth]{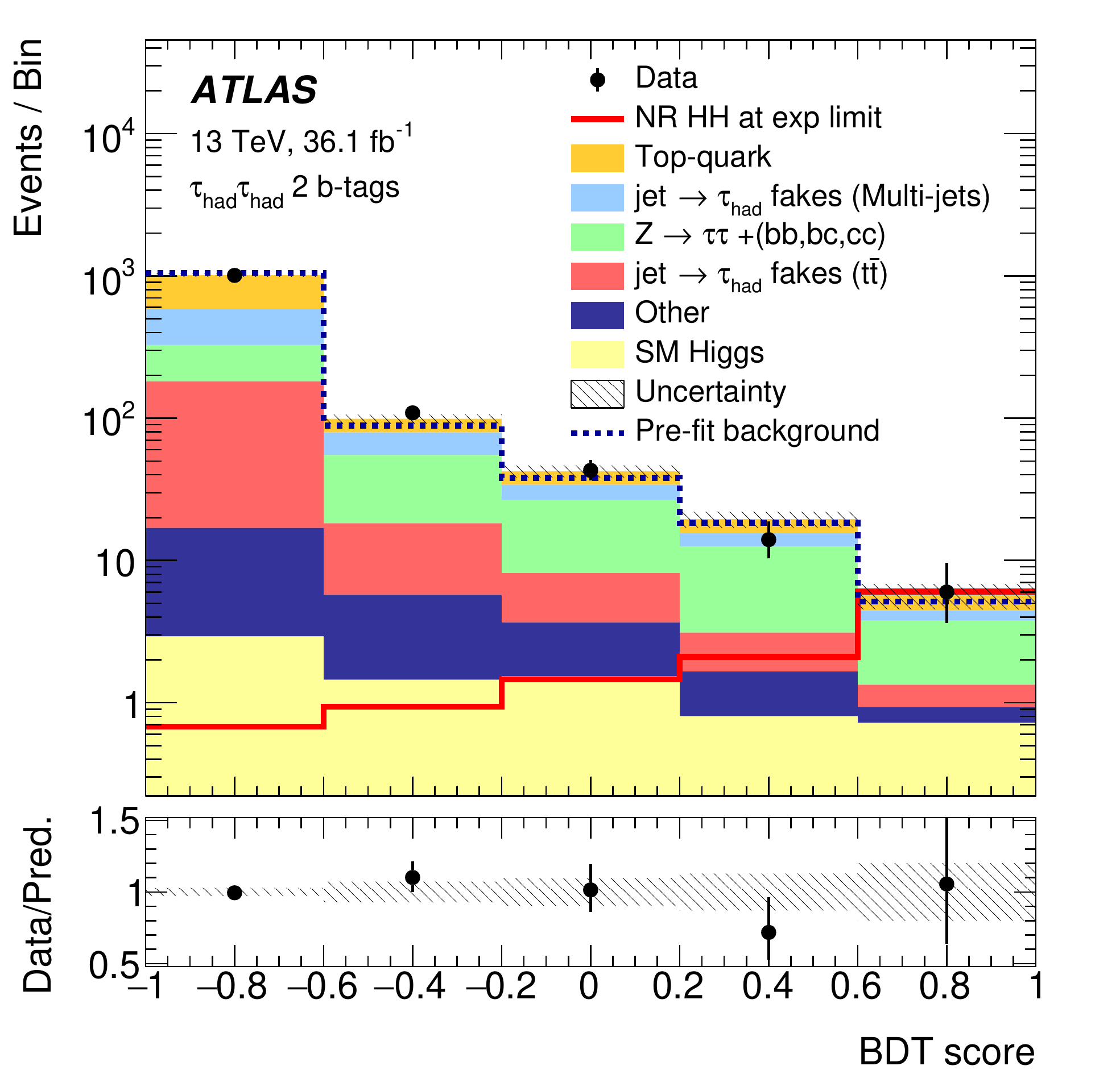}
  }
  \subfloat[]{
    \includegraphics[width=0.35\textwidth]{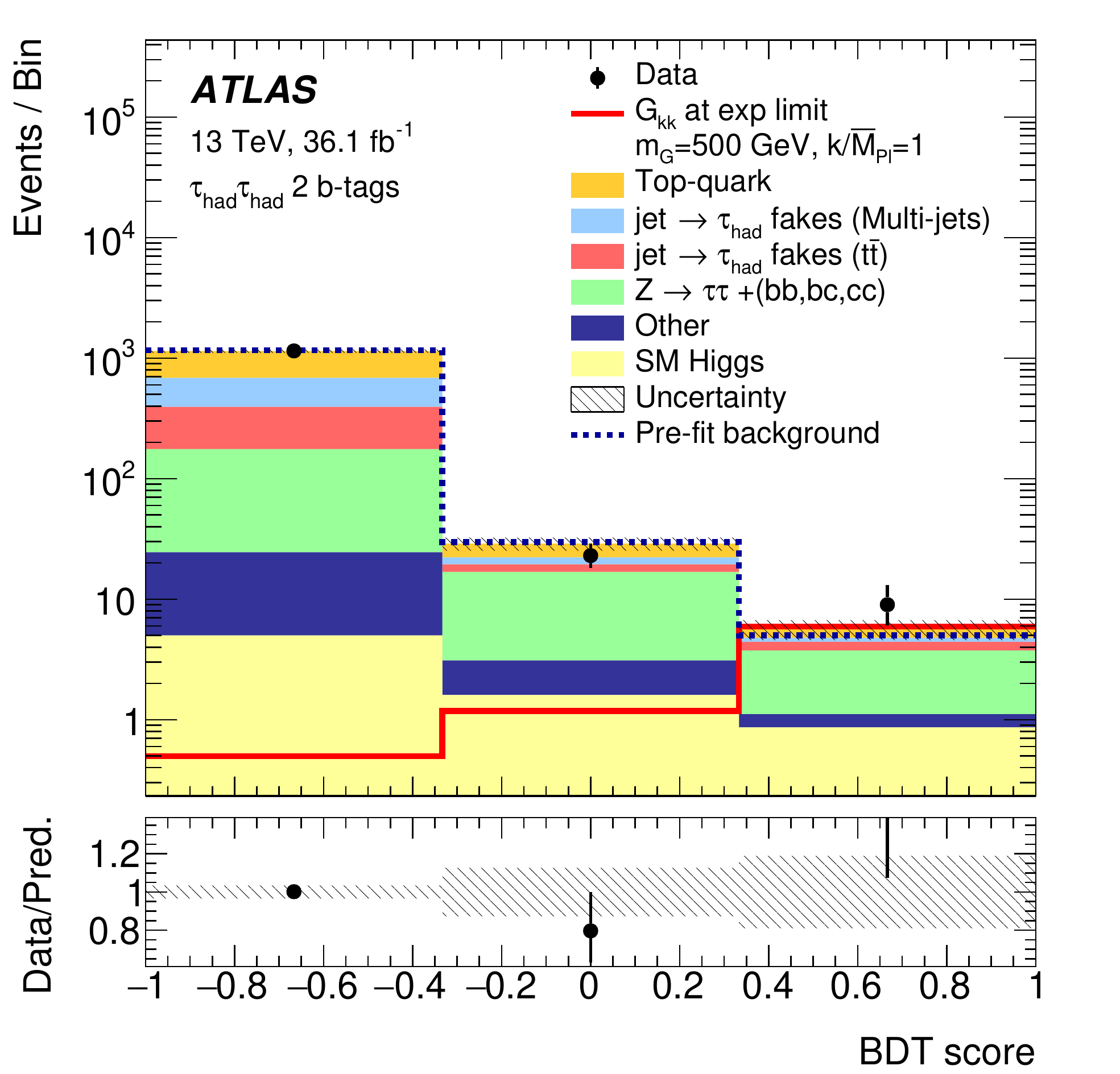}
  }

  \caption{Distributions of the BDT score for NR $HH$ signal (left) and bulk RS signal with
    $m_{G_{\mathrm{KK}}}$ = 500 GeV and $k/\overline{M}_\mathrm{Pl}$~=~1 (right) in the (a, b)
    \tlhad single-lepton trigger (SLT), (c, d) lepton + $\tauh$ trigger (LTT) and (e, f) \thadhad
    channels. Distributions are shown after the fit to the background-only hypothesis and the signal
    is scaled to approximately the expected limit.   The hatched band indicates the combined
    statistical and systematic uncertainty in the background.
    The ratio of the data to the sum of the backgrounds is shown in the lower panel.} 
  \label{fig:bdt_output}
\end{figure}

For each signal model considered, a profile-likelihood fit~\cite{Cowan:2010js} is applied to the BDT
score distributions simultaneously in the three SRs to extract the
signal cross-section, along with the \ttbar\ and $Z$+heavy-flavor normalizations.  The lattermost
is constrained by including the dedicated control region in the fit.
All sources of systematic and statistical uncertainty in the signal and background models are
implemented as deviations from the nominal model, scaled by nuisance parameters that are profiled in
the fit.  None of the dominant nuisance parameters are significantly constrained or pulled relative
to their input value by the fit.
The BDT score distributions for the non-resonant search and the $G_{\mathrm{KK}}$ signal are shown
in Figure~\ref{fig:bdt_output} after performing the fit and assuming a
background-only hypothesis. The acceptance times efficiency for the NR $HH$ signal is 3.2\% (1.9\%)
in the combined SLT and LTT \tlhad (\thadhadalt) channel over the full BDT distribution, decreasing to 1.8\% (1.4\%) for the two most
sensitive BDT bins. As no significant excess over the expected background is observed, upper limits
are set on non-resonant and resonant Higgs boson pair production at 95\% CL using the $CL_{\text{s}}$
method~\cite{CLs}.

Table~\ref{tab:results_nonresonant} presents the upper limits on the cross-section for non-resonant
$HH$ production times the $HH\rightarrow bb\tau\tau$ branching ratio, and comparisons with
the SM prediction.  The observed (expected) limit is 30.9~fb (36.0~fb), 12.7 (14.8) times the SM
prediction.  In order to compare with previous results, the BDTs are trained and applied to the signal
sample without reweighting the $m_{HH}$ spectrum to Refs.~\cite{HH_xsec,finite_mtop_2}, giving an 
observed (expected) limit of 37.4~fb (33.5~fb), 15.4 (13.8) times the SM prediction.

\begin{table}
\centering
\caption[]{Observed and expected upper limits on the production cross-section times the $HH\rightarrow bb\tau\tau$
  branching ratio for NR $HH$ at 95\% CL, and their ratios to the SM prediction.  The $\pm$1$\sigma$ variations about the expected limit are also shown.}
\label{tab:results_nonresonant}
\begin{tabular}{
    l
    l
    S[table-format=3.2]
    S[table-format=3.2]
    S[table-format=3.2]
    S[table-format=3.2]   
  }
\hline\hline
\multicolumn{2}{c}{}                                   & \text{Observed}    & \text{$-1\sigma$}  & \text{Expected}    & \text{$+1\sigma$} \\ \hline   

\multirow{ 2}{*}{\tlhad\ }& $\sigma$($HH\rightarrow bb\tau\tau$)~[fb]         &   57       & 49.9    &   69     &   96      \\
                                & $\sigma$/$\sigma_{\mathrm{SM}}$             &   23.5     & 20.5    &   28.4     &   39.5      \\ \hline

\multirow{ 2}{*}{\thadhad\ }   & $\sigma$($HH\rightarrow bb\tau\tau$)~[fb]  &  40.0      & 30.6    & 42.4       &   59      \\
                                & $\sigma$/$\sigma_{\mathrm{SM}}$             &  16.4      & 12.5    &   17.4     &   24.2      \\ \hline

\hline 
\multirow{ 2}{*}{Combination}   & $\sigma$($HH\rightarrow bb\tau\tau$)~[fb] &  30.9      & 26.0    &   36.1     &   50      \\
                                & $\sigma$/$\sigma_{\mathrm{SM}}$             &  12.7      & 10.7    &   14.8     &   20.6      \\ \hline\hline

\end{tabular}
\end{table}

\begin{figure}
\begin{minipage}{0.5\textwidth}
    \includegraphics[width=1.0\textwidth]{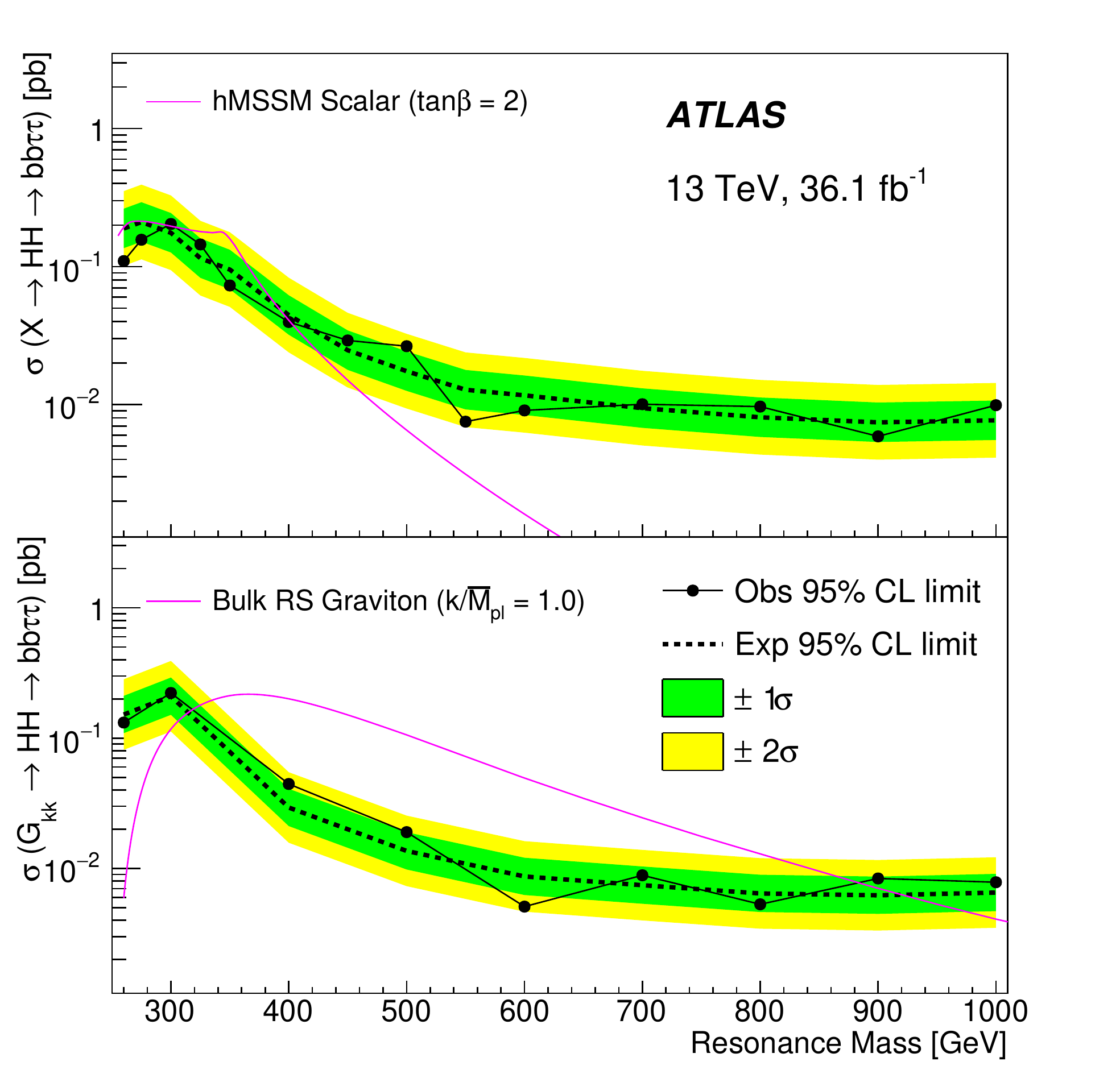}
  \end{minipage}
\begin{minipage}[b]{0.5\textwidth}
  \caption{Observed and expected limits at 95\% CL on the cross-sections of a generic narrow-width scalar $X$ (top) and RS
    $G_{\mathrm{KK}}$\ (bottom) times the branching fraction to two $CP$-even Higgs bosons $H$, when
    combining the \tlhad and \thadhad channels. The expected cross-section for the hMSSM scalar $X$
    production at $\tan\beta=2$ and the bulk RS graviton production with $k/\overline{M}_\mathrm{Pl}$ = 1.0 are also shown in the respective plots. In the hMSSM case, the bump in the theory prediction around 350 GeV corresponds to the threshold for $X$ decaying into \ttbar\ pairs.}
  \label{fig:limits}
\end{minipage}
\end{figure}

The results of searches for resonant $HH$ production are presented as exclusion limits
on the cross-section times the $HH \to bb\tau\tau$ branching ratio as a function of the resonance mass. The
expected and observed limits for narrow-width scalar resonances $X$ and $G_{\mathrm{KK}}$ signal models are shown in
Figure~\ref{fig:limits}.
For scalar resonances, the results are interpreted in a simplified minimal supersymmetric model, the
hMSSM~\cite{Djouadi:2013vqa,Djouadi:2013uqa}, where the mass of the light $CP$-even Higgs boson is
fixed to 125 GeV.  The mass range 305~\GeV\ < $m_X$ < 402~\GeV\ is excluded at 95\%~CL for
$\tan\beta$~=~2, where $\tan\beta$ is the ratio of the vacuum expectation values of the scalar
doublets.  Gravitons are excluded at 95\%~CL in the mass range 325~\GeV\ < $m_{G_{\mathrm{KK}}}$\ <
885~\GeV\ assuming $k/\overline{M}_\mathrm{Pl}=1$.  Above $\sim 600$~\GeV, the limits are largely
insensitive to the value of $k/\overline{M}_\mathrm{Pl}$, while at low $m_{HH}$ they improve
significantly with increasing $k$ due to the larger natural width.  The limits on resonant $HH$ production are
significantly more stringent than previous results in the $bb\tau\tau$ channel and competitive with
limits obtained in other channels.

\FloatBarrier

In summary, a search for resonant and non-resonant Higgs boson pair production in the $bb\tau\tau$
final state is conducted with 36.1 fb$^{-1}$ of $pp$ collision data delivered by the LHC at
$\sqrt{s}=13$~\TeV\ and recorded by the ATLAS detector. The analysis of non-resonant Higgs pair
production excludes an enhancement of the SM expectation by more than a factor of 12.7 at 95\% CL.
This is the most stringent limit on $HH$ production to date.
Upper limits are set on resonant Higgs boson pair production for a narrow-width scalar $X$ and a spin-2 Kaluza--Klein graviton $G_\mathrm{KK}$ in the bulk RS model.

\section*{Acknowledgments}



We thank CERN for the very successful operation of the LHC, as well as the
support staff from our institutions without whom ATLAS could not be
operated efficiently.

We acknowledge the support of ANPCyT, Argentina; YerPhI, Armenia; ARC, Australia; BMWFW and FWF, Austria; ANAS, Azerbaijan; SSTC, Belarus; CNPq and FAPESP, Brazil; NSERC, NRC and CFI, Canada; CERN; CONICYT, Chile; CAS, MOST and NSFC, China; COLCIENCIAS, Colombia; MSMT CR, MPO CR and VSC CR, Czech Republic; DNRF and DNSRC, Denmark; IN2P3-CNRS, CEA-DRF/IRFU, France; SRNSFG, Georgia; BMBF, HGF, and MPG, Germany; GSRT, Greece; RGC, Hong Kong SAR, China; ISF, I-CORE and Benoziyo Center, Israel; INFN, Italy; MEXT and JSPS, Japan; CNRST, Morocco; NWO, Netherlands; RCN, Norway; MNiSW and NCN, Poland; FCT, Portugal; MNE/IFA, Romania; MES of Russia and NRC KI, Russian Federation; JINR; MESTD, Serbia; MSSR, Slovakia; ARRS and MIZ\v{S}, Slovenia; DST/NRF, South Africa; MINECO, Spain; SRC and Wallenberg Foundation, Sweden; SERI, SNSF and Cantons of Bern and Geneva, Switzerland; MOST, Taiwan; TAEK, Turkey; STFC, United Kingdom; DOE and NSF, United States of America. In addition, individual groups and members have received support from BCKDF, the Canada Council, CANARIE, CRC, Compute Canada, FQRNT, and the Ontario Innovation Trust, Canada; EPLANET, ERC, ERDF, FP7, Horizon 2020 and Marie Sk{\l}odowska-Curie Actions, European Union; Investissements d'Avenir Labex and Idex, ANR, R{\'e}gion Auvergne and Fondation Partager le Savoir, France; DFG and AvH Foundation, Germany; Herakleitos, Thales and Aristeia programmes co-financed by EU-ESF and the Greek NSRF; BSF, GIF and Minerva, Israel; BRF, Norway; CERCA Programme Generalitat de Catalunya, Generalitat Valenciana, Spain; the Royal Society and Leverhulme Trust, United Kingdom.

The crucial computing support from all WLCG partners is acknowledged gratefully, in particular from CERN, the ATLAS Tier-1 facilities at TRIUMF (Canada), NDGF (Denmark, Norway, Sweden), CC-IN2P3 (France), KIT/GridKA (Germany), INFN-CNAF (Italy), NL-T1 (Netherlands), PIC (Spain), ASGC (Taiwan), RAL (UK) and BNL (USA), the Tier-2 facilities worldwide and large non-WLCG resource providers. Major contributors of computing resources are listed in Ref.~\cite{ATL-GEN-PUB-2016-002}.








\printbibliography

\clearpage
 
\begin{flushleft}
{\Large The ATLAS Collaboration}

\bigskip

M.~Aaboud$^\textrm{\scriptsize 34d}$,    
G.~Aad$^\textrm{\scriptsize 99}$,    
B.~Abbott$^\textrm{\scriptsize 124}$,    
O.~Abdinov$^\textrm{\scriptsize 13,*}$,    
B.~Abeloos$^\textrm{\scriptsize 128}$,    
D.K.~Abhayasinghe$^\textrm{\scriptsize 91}$,    
S.H.~Abidi$^\textrm{\scriptsize 164}$,    
O.S.~AbouZeid$^\textrm{\scriptsize 39}$,    
N.L.~Abraham$^\textrm{\scriptsize 153}$,    
H.~Abramowicz$^\textrm{\scriptsize 158}$
% ...
\bigskip
\\

% ...
$^{13}$Institute of Physics, Azerbaijan Academy of Sciences, Baku; Azerbaijan.\\
% ...
$^{34}$$^{(a)}$Facult\'e des Sciences Ain Chock, R\'eseau Universitaire de Physique des Hautes Energies - Universit\'e Hassan II, Casablanca;$^{(b)}$Centre National de l'Energie des Sciences Techniques Nucleaires (CNESTEN), Rabat;$^{(c)}$Facult\'e des Sciences Semlalia, Universit\'e Cadi Ayyad, LPHEA-Marrakech;$^{(d)}$Facult\'e des Sciences, Universit\'e Mohamed Premier and LPTPM, Oujda;$^{(e)}$Facult\'e des sciences, Universit\'e Mohammed V, Rabat; Morocco.\\
% ...
$^{39}$Niels Bohr Institute, University of Copenhagen, Copenhagen; Denmark.\\
% ...
$^{91}$Department of Physics, Royal Holloway University of London, Egham; United Kingdom.\\
% ...
$^{99}$CPPM, Aix-Marseille Universit\'e, CNRS/IN2P3, Marseille; France.\\
% ...
$^{124}$Homer L. Dodge Department of Physics and Astronomy, University of Oklahoma, Norman OK; United States of America.\\
% ...
$^{128}$LAL, Universit\'e Paris-Sud, CNRS/IN2P3, Universit\'e Paris-Saclay, Orsay; France.\\
% ...
$^{153}$Department of Physics and Astronomy, University of Sussex, Brighton; United Kingdom.\\
% ...
$^{158}$Raymond and Beverly Sackler School of Physics and Astronomy, Tel Aviv University, Tel Aviv; Israel.\\
% ...
$^{164}$Department of Physics, University of Toronto, Toronto ON; Canada.\\
% ...
$^{a}$ Also at Borough of Manhattan Community College, City University of New York, NY; United States of America.\\
$^{b}$ Also at Centre for High Performance Computing, CSIR Campus, Rosebank, Cape Town; South Africa.\\
$^{c}$ Also at CERN, Geneva; Switzerland.\\
$^{d}$ Also at CPPM, Aix-Marseille Universit\'e, CNRS/IN2P3, Marseille; France.\\
$^{e}$ Also at D\'epartement de Physique Nucl\'eaire et Corpusculaire, Universit\'e de Gen\`eve, Gen\`eve; Switzerland.\\
% ...
$^{*}$ Deceased

\end{flushleft}



\end{document}